\documentclass[10pt]{article}
\usepackage[parfill]{parskip}
\usepackage{geometry}
\usepackage{enumerate}
\usepackage{graphicx}
\usepackage{mathtools}
\usepackage{listings}
\usepackage{url}
\usepackage{authblk}
\usepackage{hyperref}
\usepackage{amsfonts}
\usepackage{amsmath,amssymb}
\usepackage{xcolor}
\begin{document}
\date{}
\title{Indexing structures for the PLS blockchain}
\author{Alex Shafarenko\footnote{Correspondence address: University of Hertfordshire, U.K.; 
email: \href{a.shafarenko@herts.ac.uk}{a.shafarenko@herts.ac.uk}}}
\affil{University of Hertfordshire, AL10 9AB, UK; Ada Finex Ltd}
\maketitle

\def\things{{\it things}}
\def\thing{{\it thing}}

\eject
\begin{abstract}
This paper studies known indexing structures from a new point of view: minimisation of data exchange between an IoT device acting as a blockchain client and the blockchain server running a protocol suite that includes two Guy Fawkes protocols, PLS and SLVP. The PLS blockchain is not a cryptocurrency instrument; it is an immutable ledger offering guaranteed non-repudiation to low-power clients {\em without} use of public key crypto.  The novelty of the situation is in the fact that every PLS client has to obtain a proof of absence in all blocks of the chain to which its counterparty does not contribute, and we show that it is possible without traversing the block's Merkle tree. 

We obtain weight statistics of a leaf path on a sparse Merkle tree theoretically, as our ground case. Using the theory we quantify the communication cost of a client interacting with the blockchain. We show that large savings can be achieved by providing a bitmap index of the tree compressed using Tunstall's method. We further show that even in the case of correlated access, as in two IoT devices posting messages for each other in consecutive blocks, it is possible to prevent compression degradation by re-randomising the IDs using a pseudorandom bijective function. We propose a low-cost function of this kind and evaluate its quality by simulation, using the avalanche criterion.    

{\bf keywords:} PLS blockchain, Guy Fawkes protocol, content-addressable storage, data–structure statistics, Tunstall coding, pseudorandom bijections
\end{abstract}

\section{Introduction\label{intro}}

This paper gives statistical analysis of some known data structures required for the implementation of the PLS (permissioned) blockhain \cite{PLS} or PLSB for short, whose purpose is to support a {\em swarm} of IoT devices, or \things\ operating on the premises of a single administrative authority, for example a smart hospital. The use of a blockchain is for the purposes of audit trail, authentication and non-repudiation of {\bf all} actors, both human and unmanned, including small, bare-metal microcontrollers that supply critical sensor data and those which drive actuators. 

The utility of permissioned distributed ledger systems (permissioned blockchains, or PBCs for short) is based on two fundamentals: (i) distributed validity check of messages and (ii) an immutable, linearly-ordered ledger. In IoT applications, especially in sensor-networks, (ii) tends to be more important than (i). Indeed, typically messages are not transactions in the financial sense, so checks such as double spending are not relevant; value checks are domain-specific and are best performed by smart contracts, which leaves the authenticity and provenance of each message posted on the ledger as the only general validity concerns. The PLS blockchain \cite{PLS} assures (ii) by employing Guy-Fawkes Protocols (GFPs) \cite{GFP}. 

A GFP is a post-quantum signature protocol based on an unlimited series of interlocking cryptographic hashes. GFP computations are fast, messages short and secrets  neither moved nor kept for a long time; the GFPs are resistant to quantum computing as they do not use operations such as prime-number factorisation or discrete logarithm. Finally, by their recursive nature, GFPs define a single sequence of signatures that is very hard to split; this makes them quite suitable as a basis of a blockchain. 

In the next section we will briefly outline the architecture and protocols of the PLS blockchain published in an earlier paper\cite{PLS}. Operational differences between the PLS and other blockchains, such as Etherium, call for re-evaluation of major data structures required for its implementation. In Section \ref{sec:motivation} we argue that the limited number of users (IoT devices on the premises and human actors\footnote{In the sequel, when it is not important what kind of actor is meant we will call all actors users for short and apply the pronoun 'it'.}) and the limit on their communication duty cycle and disposable energy need efficient secure data structures to avoid communicating irrelevant data. We propose a Merkle-Tunstall Tree for that purpose (Section \ref{sec:MTT}) and provide a statistical evaluation of its efficiency. The efficiency of the Tunstall compressor depends on the lack of correlations between different user's contributing to the same block. To decorrelate block access we propose to use a random permutation function to map users' true IDs onto local IDs for a given block, see Section \ref{sec:rand}. To illustrate how proposed technologies work together we give one illustrated example in section \ref{sec:together}. Finally, there is a section on related work and some conclusions.

\paragraph{The main contributions of the paper are as follows:}

\begin{enumerate}
\item {\em Statistical analysis of a sparse Merkle tree with uniform, uncorrelated probabilistic leaf occupancy.} We have obtained the path-weight probability distribution function (as a recurrence relation in the tree height) analytically, without Monte-Carlo simulation. It is easy to quantify the function numerically for any given height. 
\item {\em The proposal and evaluation of a compressed bitmap and local enumeration of block users.} This makes it possible for a user to obtain the proof of absence in the block directly from the broadcast root of trust without accessing the Merkle tree. We have also shown that the local enumeration results in a path weight similar to that on the original sparse Merkle Tree on average, but the variation is tightly bounded from above, which makes it possible to limit the packet length when communicating a secure leaf path using our structure. By contrast, the path across the original tree varies more widely depending on the leaf statistics and may result in paths exceeding the maximum packet size.  
\item {\em The proposal and evaluation of shift-shuffle} as a low-cost pseudorandom permutation technique sufficient to break a possible correlation between occurrences of different users' records in block contents. We quantified the number of rounds in the permutation algorithm to be used taking the avalanche criterion as a basis. 
\end{enumerate}

\section{PLS blockchain: architecture and protocols.}

\begin{figure}
\begin{center}
\includegraphics[width=0.8\textwidth]{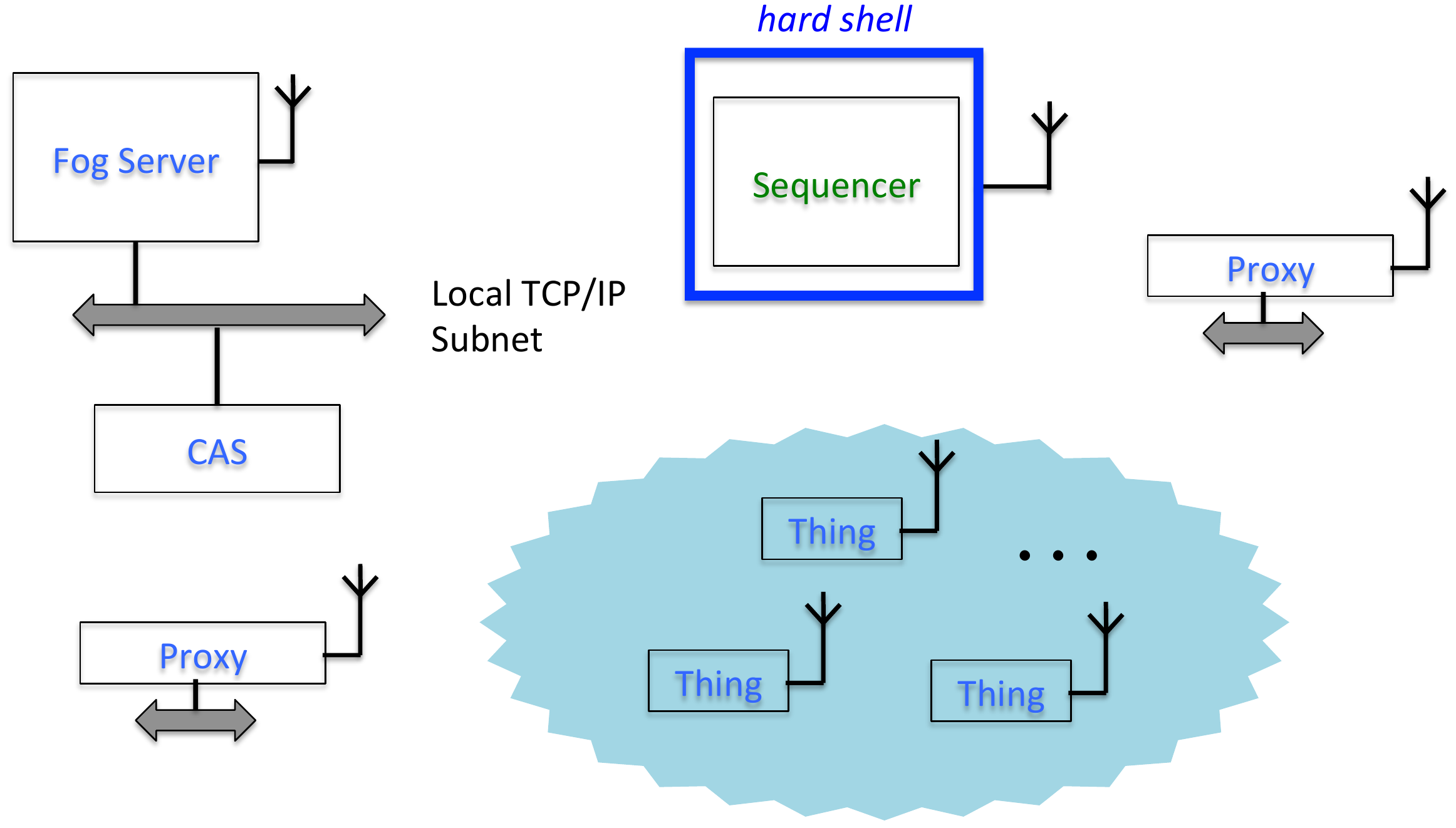}
\end{center}
\caption{Architecture of a PLS-blockchain system\label{fig:arch}}
\end{figure}

The details of the protocols and their security analysis are available from \cite{PLS}. We present them here for completeness. However, for the contributions of the present paper we only need to discuss the logistics of the PLSB, whose architecture is shown in figure \ref{fig:arch}. 

\paragraph{Blocks} are formed from transactions communicated by \things\ via {\em proxies} that make it possible for all \things\ to rely on low-power radio communication. To authorise a transaction, \things\ run another GFP protocol, called SLVP. That protocol's messages are forwarded by one or more proxies to the Fog Server (FS) to be included in the next block. The FS forms blocks regularly, on a fixed wall-clock schedule, by validating incoming SLVP messages from \things, and adding them to the current block. 

\begin{figure}
\begin{center}
\includegraphics[width=0.8\textwidth]{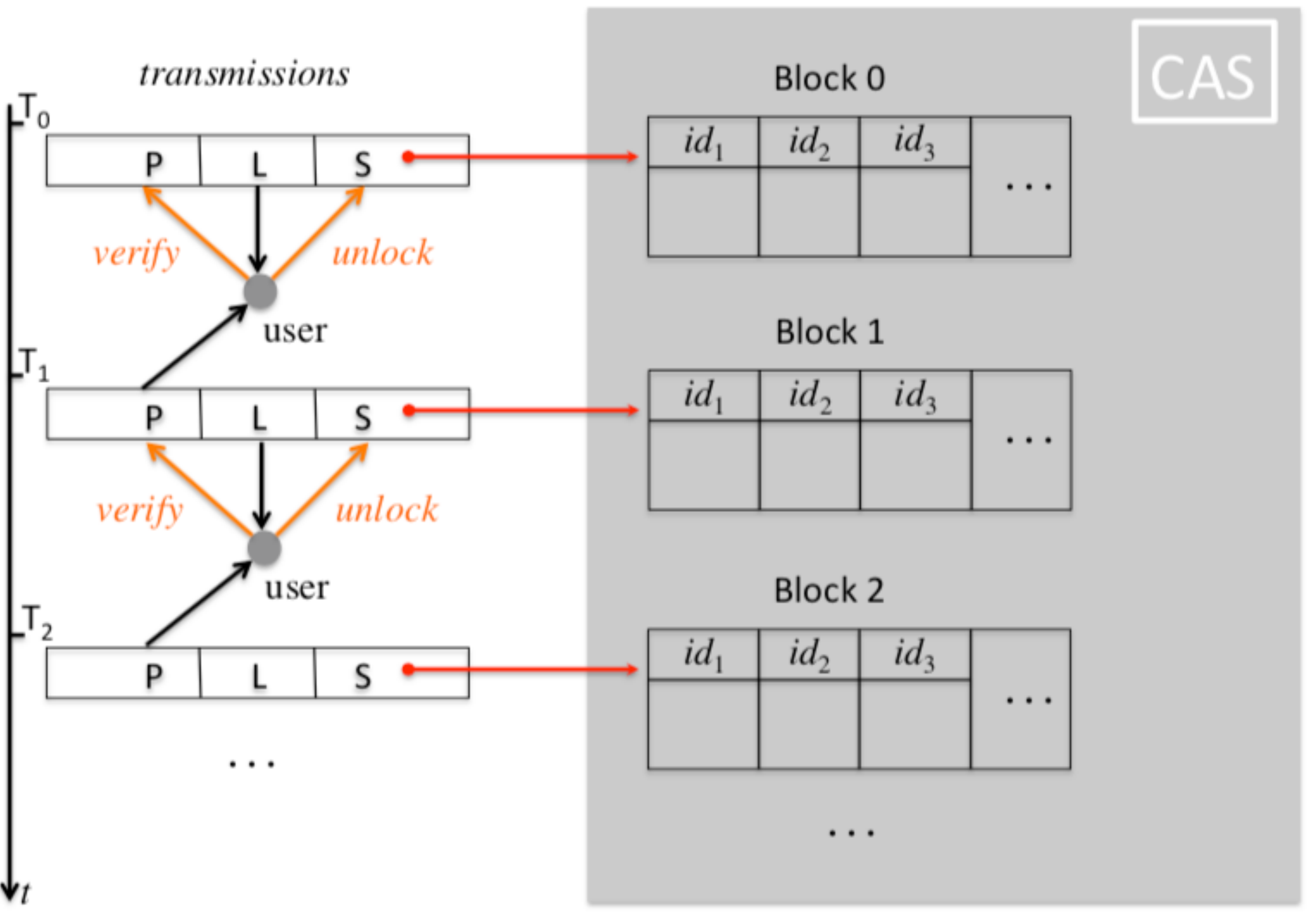}
\end{center}
\vskip1em
\hskip-0.25in\bgroup
\def\arraystretch{1.3}%
\begin{tabular}{|c|l|l|l|}
\hline
 Interval& Transmit/Receive & Verify & Unlock\\
\hline
\hline
$[T_0,T_1]$&$L_0=H(N_{1}) \oplus N_0$               &&   \\
 &$S_0={\bf E}_{N_0} (J_0\oplus H(N_{1}))$ & $P_0$ out of band &~ \\
 &$P_0=H(N_0)$                            &                           &\\

\hline

$[T_1,T_2]$&$L_1=H(N_{2}) \oplus N_1$               &                        &  \\
 &$S_1={\bf E}_{N_1} (J_1\oplus H(N_{2}))$ & $H(L_0\oplus P_1)=P_0$ & $H(B_0) = P_1 \oplus {\bf D}_{L_0\oplus P_1}(S_0)$\\
 &$P_1=H(N_1)$                            &                        & \\

\hline
...&...&...&...\\

\hline

$[T_k,T_{k+1}]$&$L_k=H(N_{k+1}) \oplus N_k$             &                                &  \\
 &$S_k={\bf E}_{N_k} (J_k\oplus H(N_{k+1}))$ & $H(L_{k-1}\oplus P_k)=P_{k-1}$ &$H(B_{k-1}) = P_k \oplus {\bf D}_{L_{k-1}\oplus P_k}(S_{k-1})$\\
 &$P_k=H(N_k)$                              &                                &\\

\hline
\end{tabular}
\egroup
\vskip1em
Notation: \\*[1em]
\hangindent=1em 
$H(\cdot)$: a cryptographic hash function\\ 
$J_i$: a digest of blockchain block $B_i$, e.g. $H(B_i)$\\
${\bf E}_\alpha$: encipherment under symmetric key $\alpha$; use PCBC mode if $J$ exceeds cipher block size\\
${\bf D}_\alpha$: decipherment under symmetric key $\alpha$, matching ${\bf E}_\alpha$\\

\caption{Structure of the PLS protocol\label{fig:pls-prot}}
\end{figure}

\paragraph{Chain.} By regular deadlines the current blocks are stored in CAS and their hash is signed by producing messages of the main protocol, PLS. All PLS messages are generated and transmitted by radio via a sealed unit, Sequencer, which receives the current block's hash from the FS on a private radio channel. The Sequencer does not contain a changeable program and is not connected to the Internet, so it is not hackable. The PLS sequence, i.e. the sequence of PLS messages, requires a short-term secret, which is produced inside the Sequencer using a physical source of randomness in one time interval  and is revealed in the next interval at the same time as selecting a new random secret. All \things\ must receive each PLS message, validate it, and unlock the corresponding block's hash, which is a file name of the block in CAS, see fig \ref{fig:pls-prot}. P- and L- messages cross-validate as shown in the figure, and S-messages contain some redundancy, which, after deciphering, indicates whether the message is valid or not. For example, $J$ can include a run of zeros at the end; this would be sufficient to thwart a ``random message'' attack, which is a possible DoS action of the attacker jamming the radio channel\footnote{Since the attacker does not know the preimage of $P$ at the time when an attack is possible, it can only send an arbitrary message; after unlocking, it would produce a near-random bitstring as a would-be block hash. The requirement for it to have $r$ trailing zeros will only be satisfied with the probability $2^{-r}$}. Also notice that blocks of the blockchain are, as usual, key-value collections, where the key is the originator's ID. 

Any invalid messages, possibly sent by an attacker will fail the validity check with a very high probability. Progress is assured by limiting the number of invalid messages using various techniques discussed in \cite{PLS}, but those are exclusively DoS countermeasures which do not influence the semantics of the blockchain. The initial message $P_0$ is authenticated by all blockchain users via external credentials. Users joining the system later would require external authentication of the latest P-message instead of $P_0$. The unlocked hashes of all the subsequent blocks are as secure as the weaker of the credential and the computational hardness of the full hash preimage problem (i.e. finding {\em all} bit-strings of a given length whose hash is a given value). The latter is not feasible for a SHA-2 hash even Post Quantum. Also notice that the verification and unlocking computations are fast (single microseconds) even for a small bare-metal microcontroller equipped with a crypto-accelerator, e.g. ESP32\cite{ESP32}.

\paragraph{Transactions.} As mentioned earlier, a \thing\ publishes a transaction on the blockchain by running the SLVP protocol with the FS. A transaction requires one round of the SLVP protocol, which takes three blockchain blocks. For the security of the protocol it is required that the originating \thing\ check that the latest sent SLVP message has appeared on a block. As soon as it has, the next protocol message can be sent. The first message to send is an  S-message, which contains the data object to be signed. Then an LV-message is posted on the blockchain, which provides interlocking hashes and verification data (the latter is needed to thwart jam-spoof attacks, see \cite{PLS}). Finally, the \thing\ posts a (proof) P-message. The FS validates the P-message using the data contained in the previous round's P-message and the content of the LV-message sent in between. The FS will only include a P-message in a block if the P-message is valid, while LV- and S- are posted right away, the reason being that invalid LV- or S-messages will be recognised as such by the protocol itself only after the next P-message is posted. In practice The FS and a user may share a secret to help the FS to authenticate incoming messages early to make it difficult for an attacker to post a large number of invalid S- and LV-messages. However, this does not help a counterparty that must be mistrustful of the FS. So additional authentication, if present, is purely a DoS countermeasure; we needn't focus on it as we concern ourselves only with the machinery of the blockchain. 

\begin{figure}
\vskip-1in
\begin{center}
\includegraphics[width=0.8\textwidth]{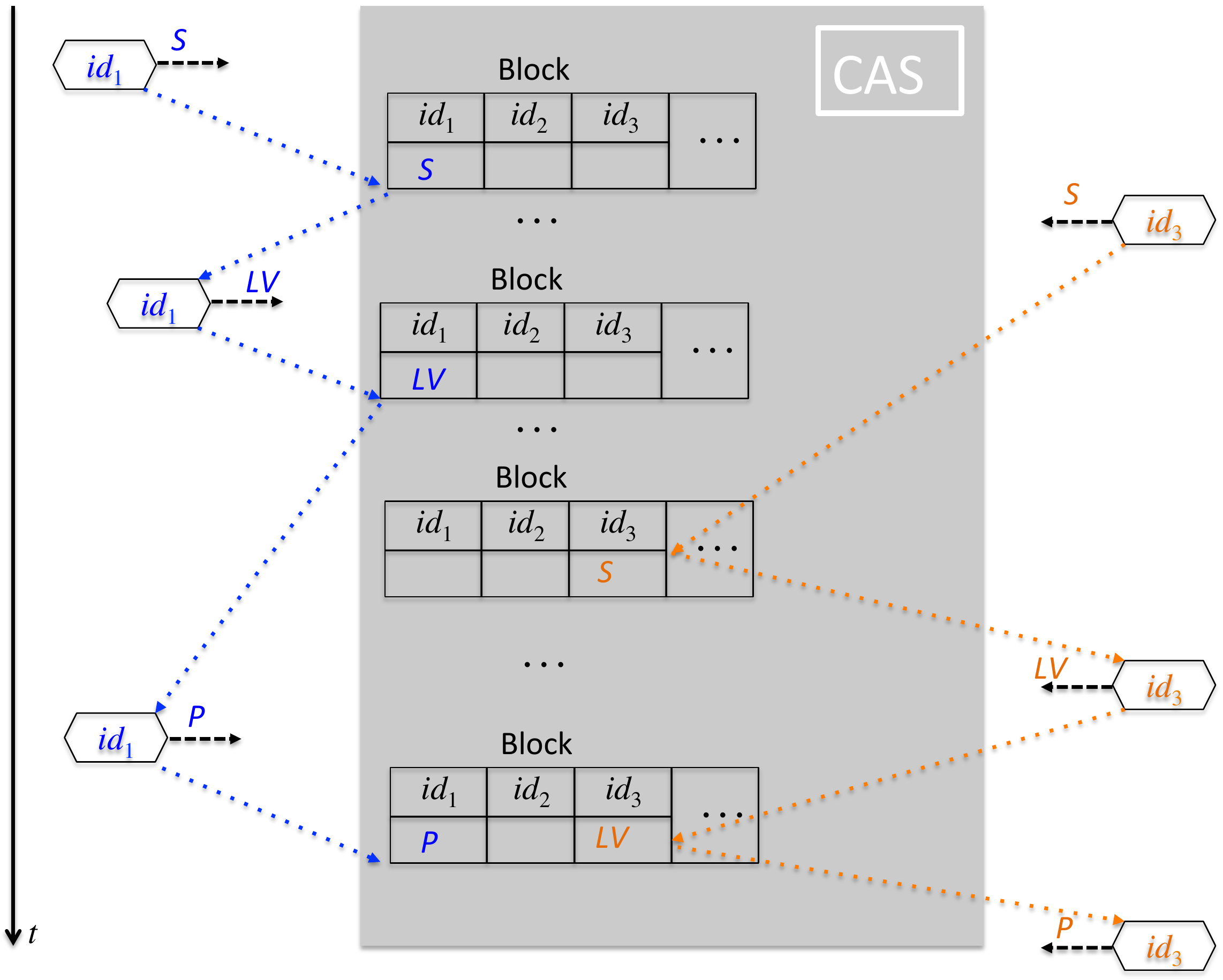}
\end{center}
\vskip1em
\bgroup
\def\arraystretch{1.3}%
\begin{tabular}{|c|l|l|l|}
\hline
 \bf Block& \bf Transmit & \bf Verify & \bf BC Action\\
\hline
\hline
$b_0$ &$P_1=H(N_1)$                                    &   Out of Band (Enrolment)                   & Post\\
\hline
$b_1$& $S_1={\bf E}^*_{N_1} (H(N_{2}), M_1)$ &   ---     & Post \\
\hline
$b_2$& $LV_1=H(N_{2}) \oplus N_1\,||\, H(H(N_2)||N_1)$         &     ---                 & Post\\
\hline
\hline

...&...&...&...\\

\hline
\hline
$b_{n}$ &$P_{k}=H(N_{k})$  &{\em as for $b_{n+3}$, assume success}& Post\\
\hline
$b_{n+1}$& $S_{k}={\bf E}^*_{N_{k}}(H(N_{k+1}), M_{k})$ &    ---        & Post  \\
\hline
$b_{n+2}$& $LV_{k}=H(N_{k+1}) \oplus N_{k}\,||\, H(H(N_{k+1})||N_{k})$ &        ---                 & Post\\
\hline
\hline
$b_{n+3}$ &$P_{k+1}=H(N_{k+1})$  & for $b_n<b<b_{n+3}$:&\\
~&~&\hskip1em find {\bf first} LV such that &\\ 
~&~&\hskip2em $H(P_{k+1}||L\oplus P_{k+1})=V$ &\\ 
~&~&{\bf if found in block} $\hat{b}$:&  \\ 
~&~&\hskip1em{\bf unless} $\exists b\in(b_n,\hat{b}), L^\prime\in B_b:$&\\
~&~&\hskip2em$H(L^\prime\oplus P_{k+1})=P_k$& Post\\
~&~&{\bf else}& Reject \\
\hline
\hline
\end{tabular}
\egroup
\vskip1em
To unlock the data object $M_k$ compute $M={\bf D}^*_{L\oplus P_{k+1}}(P_{k+1},S)$ for every S-message in the interval $[b_n,\hat{b})$ and accept the first valid $M$. Here $E^*_k(a,b)$ is the encryption of $b$ under key $k$ in PCBC mode with $a$ as IV; $D^*_k(a,b)$ is the matching decryption.

\caption{Structure of the SLVP protocol. Table on behalf of a single user\label{fig:slvp-prot}}
\end{figure}

The protocol is summarised in figure \ref{fig:slvp-prot}. The diagram shows two users, blue and brown, posting their SLVP messages on the blockchain using the transmissions shown in the table below, which is presented on behalf of a single blockchain user. Users are independent in transmitting protocol messages for themselves and in verifying messages sent by  others. The verification formulae in the penultimate column enable users (as well as the FS that does it first) to prove to themselves that the other party has genuinely signed its data object $M$. If the next P-message, $P_{k+1}$ checks out, they use its value to unlock the data object $M_k$ as defined in the bottom of the figure\footnote{The original paper \cite{PLS} has a slightly different arrangement for S-messages since in the original design CAS was trusted for progress, but in the present paper we eliminate this requirement.}. Just like blockchain blocks, the data objects have some redundancy. For example, a certain number of leading zeroes, not necessarily very few as this does not require computation, will adequately defend against random S messages, sent by an attacker. Notice that the SLVP protocol defines {\em variable length} encryption for S-messages using a block cipher in PCBC mode. Encryption is bijective, i.e. information-preserving, and the redundancy required for validation in the presence of a random-message attack is just a few bytes (e.g. 4 bytes gives the attacker 1 chance in a few billion to post valid random data, but even then it only subverts a single S-message). 

Also note that the FS has authority to introduce a new user by posting their very first P-message. The first P-message is always marked as such on the blockchain for other users exchanging transactions to recognise it as the originator's identity.

\paragraph{Operations.} Transactions can be posted by both \things\ and human users. Each \thing\ has one or more {\em masters}, which are typically users (but could be other \things). Not only does each \thing\ check the posting of each of its SLVP messages on the blockchain, it also monitors the postings of all its masters and any relevant counterparties, and validates their data objects by applying the SLVP protocol. Alternatively, the \thing\ can participate in a smart contract which would only require it to follow and validate messages from the contract engine. In this paper we limit ourselves to the mechanics of transaction processing, while leaving higher-level protocol to further work. We will assume in the sequel that each \thing\ is interacting with a very small number of other actors and needs to follow a few SLVP threads (perhaps 2 or 3). Our focus will be on how to make these interactions as computationally and communicationally efficient as possible. 

\paragraph{Addresses.} Each PLS user has an address, which is a small number. Since we concern ourselves with a localised enterprise solution (e.g. a smart hospital) covered by a direct link radio network (e.g. smart sensors equipped with a LoRa\cite{LoRa} transceiver), we do not expect the number of \things\ greater than circa 1000. The total number of actors should be a small factor of that to account for human users and smart contracts, so 2--4K addresses is our target. Transactions have an originating address and a destination. 

\paragraph{Frequency.} In IoT networks of interest, communication is limited by the duty cycle to save the limited bandwidth that all \things\ have to share. This is in addition to the constraints imposed by the energy budget of an individual IoT device. Consequently a small fraction (typically a few percent) of the registered users will be posting a transaction in any given block.  

\section{Block structure and optimisation challenge} 

\paragraph{Immutable dictionary.} In the previous section blockchain blocks were shown in the diagrams that consisted of records attributed to various users as key-value pairs, where the key is the user ID. In a given block only some of the registered users would be represented by records. Since the frequency of posting on the blockchain in our case is severely limited by the \things' communication duty cycle (if using LoRa) or energy budget (if using public networks or LoRa), the proportion of users posting to any given block is expected to be very small. Still, the user can only authenticate the block by the S-message of the PLS protocol, which, when unlocked, contains the block hash. On the other hand, as we mentioned in the previous section the user is typically interested in two or three other users' contributions, which, given a typical enterprise IoT network of a thousand \things, is still much less than the expected volume of activity in a block. Indeed, given a block production rate of 4 blocks per hour and a \thing\ data production rate of 2--4 samples per day, and bearing in mind that each data post requires 3 blocks according to the SLVP protocol, we arrive at 6 to 12 blocks per device per day and circa 100 blocks per day in total. This means that in the absence of correlated activities we should expect about 5 to 10 percent of the swarm to post in every given block. For a 1000-strong swarm, the block may contain an estimated 50 to 100 user records authenticated by a single hash. If a \thing\ wishes to access just a few of these, it would have to first read the whole block and check the hash to validate it, and then dispose of most of these records as they would be irrelevant. 

\begin{figure}
\begin{center}
\includegraphics[width=0.8\textwidth]{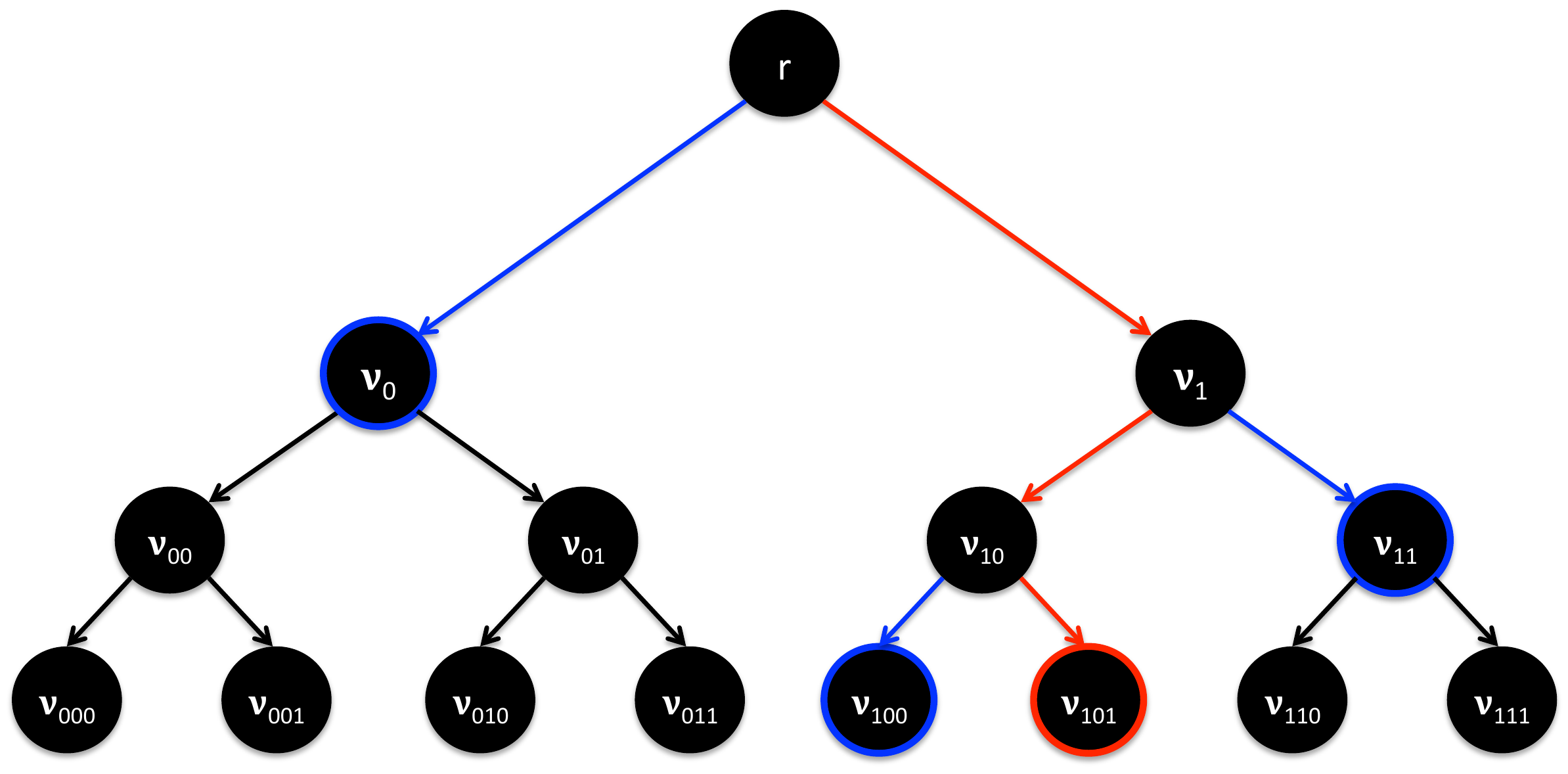}
\end{center}
\caption{Merkle Tree.\label{fig:MT}}
\end{figure}

\paragraph{Merkle tree.} The standard solution to the above problem is called the {\em Merkle Tree}(MT)\cite{MerkleTree}, see figure \ref{fig:MT}. It is a labelled binary\footnote{The tree does not have to be binary, but higher-based trees, and higher-based MPTs, discussed later, are inefficient for a small number of leaves.} tree each node $\nu$ of which has two children, with some labels $\nu_L$ and $\nu_R$, and its own label is $\nu=H(\nu_L || \nu_R)$. A child can either be a leaf or a full node in its own right; in both cases it has a label but in the latter case it also has two children of its own. It is quite clear that a change in any leaf will change the root label (also known as the {\em root hash}), so the authenticated root makes the whole tree authentic. For each node except the root there exists one other node with the same parent, which we call {\em adjunct}. What makes the MT useful is that it can also authenticate a single leaf by providing a root-path list of adjunct nodes' labels, or root-adjunct path for short. For example, to authenticate the leaf $\nu_{101}$ shown in red in the figure, given the root label $r$, we only need to know the labels of the blue (adjunct) nodes: $\nu_{100}$, $\nu_{11}$ and $\nu_0$, since 
\[
r=H(\nu_0||H(H(\nu_{100}||\nu_{101})||\nu_{11}))\,.
\]
Generally speaking, for a tree with $K$ leaves one needs to communicate $h=\left\lceil{\log_2 K}\right\rceil$ hashes, which is much less than $K$ for the number of leaves in the hundreds that we are considering. The tree thus represents an {\em array} of leaves indexed by the path: a left edge represents 0 and a right edge 1; the edges traversed en route to the leaf form a bit-string that represents the key. The leaf itself represents the value of the key-value pair. 

\paragraph{Blocks represented as Merkle trees.} Common practice in Blockchain construction is to represent a block as an MT, each leaf of which carries the hash of a user's record included in the block, with the user ID being the key gleaned from the leaf's root path. A user requesting another user's record (or the one of its own) from an intelligent CAS could just receive the root-adjunct path corresponding to the requested ID and hash it through to match with the root hash value. If the PBC signs the root hash of every block it creates, no further security is required to authenticate any user records. Our investigation is of a special case when the maximum number of users is small and is known in advance, and where good communication efficiency is important. 
We could use an MT with the tree height $h$ close to 10 (to accommodate our expected $2^{10}\sim 1000$ users). Since, as we have mentioned earlier, we expect only around 50 (maybe up to 100) users to contribute to any given block, a great majority of the leaves will not be used. 

\paragraph{Mask-controllable sparse MT.} The number of leaves in an MT does not have to be a power of 2. Also leaves can have no value associated the root-path key. We can think of such leaves as unoccupied. An MT with no-value leaves is called a {\em sparse} MT, or SMT. There are several ways of organising an SMT, but proposals usually focus on {\em mutable} trees that are used for secure updatable key-value storage. Our interest is in {\em immutable} SMTs, where efficiency is understood in narrow terms as efficiency of retrieval {\em only}. Below we define our own version of the SMT, geared towards our objectives.

We can assume that a leaf without value has a special label NULL and the parent of two NULL nodes has the label NULL as well. This assumption does not diminish security due to the fact that a NULL child of any node is implicitly associated with the node height. Consequently, the shape of the NULL subtree associated with the child is completely defined by its root position. All such NULL trees are identical anyway, so a single label value fully represents them.

For the verifier to be able to verify a path with NULL nodes, it requires a bit mask of length $h$, where bit-value 1 indicates that the corresponding adjunct node is non-NULL; and the bit-value 0, that it is NULL. The NULL labels can then be omitted from the path. Finally we extend the domain of $H(x)$ to include NULL-concatenated strings by defining that for any bit-string $x$ 
\begin{gather}
H(x\parallel{\rm NULL}) = H(x\parallel x^\prime)\,,\label{eq:none-case}\\
H({\rm NULL}\parallel x) = H(x^\prime\parallel x)\,,\nonumber
\end{gather}
where $x^\prime$ is the bit string obtained form $x$ by flipping all bits. Interestingly, a simpler extension 
\[
H(x\parallel{\rm NULL}) = H({\rm NULL}\parallel x) = H(x)
\]
would not be secure, as it allows one to construct a second preimage by rotating the subtree or swapping nodes along a NULL path. It is easy to see that the hashing process introduced by Eq \ref{eq:none-case} is not invariant to any such transformation. It is impossible to create a new valid SMT with the same root hash and a different leaf sequence without solving the second preimage problem. 

If the path is mask-controlled, CAS only needs to communicate {\bf up to $h$} hashes in addition to the bit mask for the verifier to successfully compute the root hash. Extending the above example, if $\nu_{11}$ were unoccupied, CAS could supply the bit mask 101 (the second adjunct is missing in path order), and the values of $\nu_0$ and $\nu_{100}$. Notice that the bit mask does not need to be secured: if it is incorrect, the verifier will compute an incorrect path expression and the result will not match the root hash. Also notice that the mask is very small compared to the hash length: for a tree of 1024 leaves (counting both NULL and non-NULL) the root-adjunct path contains from 1 to 10 hashes 256 bits each, i.e. 256 to 2560 bis, but the mask length is only 10 bits regardless.  

\begin{figure}
\begin{center}
\includegraphics[width=0.8\textwidth]{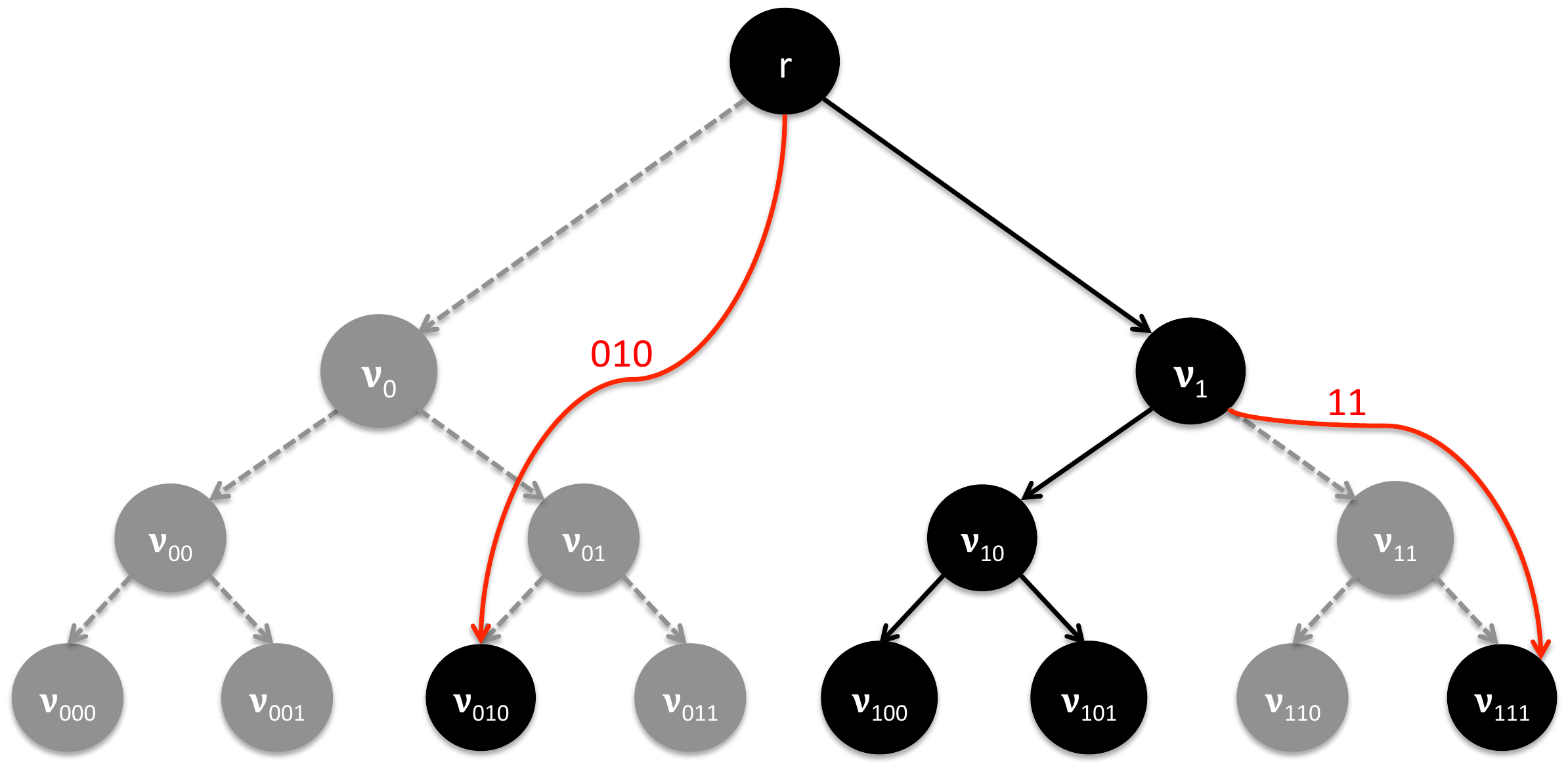}
\end{center}
\caption{Merkle-Patricia Trie. All unlabelled edges are assumed to have the label `0' if they lean to the left and `1' if they lean to the right.\label{fig:MPT}}
\end{figure}

\paragraph{Merkle-Patricia Trie.} The idea of mask-control path is similar to that of the so called {\em Merkle-Patricia Trie}(MPT)\cite{IndImmute} where not only the nodes but also the edges can be labelled. If a node has a single active edge (i.e the other edge leads to a NULL subtree), the node is eliminated and its parent uses the prefix of the other edge as its label, see figure \ref{fig:MPT}. The example in the figure is of a block where, out of the maximum 8, only users  2, 4, 5, and 7 (010, 100, 101, 111 in binary) are present. Notice that we still have a binary tree, but the root-adjunct path augmented with edge labels requires from one (for 010) to three (for 100 and 101) adjunct hashes for validation, depending on the leaf. The edge labels are typically much shorter bit-strings than a single cryptographic hash ($\log_2 K \ll 256$) and so can be neglected in determining the communication efficiency of the access scheme. The same is true of masks with our version of the SMT.

How is the edge label secured? It is simply hashed together with its child content in determining the node label: 
\[
\nu=H(\lambda_0 || \lambda_1 || \nu_0 || \nu_1)\,,
\]

where $\lambda_{0,1}$ are the edge labels of the left and right child, respectively.

It is easy to see that there is a direct correspondence between the MPT and the SMT with mask-controlled paths. Our construction requires more work when validating a path: each node, irrespective of its path quality involves $h$ hash calculations for verification, where $h$ is the height of the tree, but in the MPT case the number of times a hash is calculated is the same as the number of adjunct hashes supplied with the MPT path, although each hash calculation also involves edge labels, which may increase the cost.  The total {\em length} of edge labels along the root path in the MPT case is equal to the length of the mask in the SMT case. However, an MPT path requires markers to partition the path string into individual edge labels. Our construction is slightly more frugal in this respect, and it is simpler, which is why we prefer it. 

\section{Motivation and optimisation idea\label{sec:motivation}}

It is obvious from the SLVP protocol that an actor engaging in transactions with another on the PLSB must check each block to determine the presence of a transaction message from the counterparty. Due to the low duty-cycle of \thing-to-FS communication, the counterparty will not be present in a great majority of blocks. However, to securely establish the absence the actor must traverse the block and verify that the counterparty's record is not there. In the MT case we can use the mask-controlled path to the unoccupied leaf which can contain up to $h$ adjunct hashes. In the MPT case CAS will supply the longest path {\em in the direction of}, rather than to, the unoccupied leaf. By examining the last node on that path the user will be able to verify that the necessary edge is missing. For example, looking at the MPT in figure \ref{fig:MPT}, if an actor requested the unoccupied leaf 110, the one-step path $r\to\nu_1$ with adjunct material will be sent back for validation: 
\begin{gather}
010,1,\nu_{010}\nonumber\\
0,11,\nu_{10},\nu_{111}\nonumber
\end{gather}
The number of hashes to be communicated is the same as that for the mask-controlled MT: in the current example the latter would require hashes $\nu_{111}$, $\nu_{10}$, and $\nu_0$. An advantage of the MPT is that it saves the verifier extra hash computations by providing segments of the path as (hashed-controlled) edge labels. While saving {\em some} compute time, the effect of it is negligible, since the maximum root-adjunct path length is logarithmic in the number of leaves ($\lesssim12$), and a modern microcontroller can compute tens of thousands of hashes per second. A disadvantage of the MPT is that it requires communication of edge labels in addition to the hash for each node on the path, but again, compared to the hash length this is negligible, too. 

What is considerably more important here is that neither mask-controlled MT, nor MPT reduce the maximum root-adjunct path length. In our example the number of leaves present is 4, but the hight of the MPT as a binary tree is 3, not $\log_2 4 = 2$; it is as if all leaves were present. As a result, the system would require to accommodate longer communication packets, which may affect the guaranteed duty cycle limit of an IoT device.

This brings us to the {\bf central idea} of the paper: to index a block, one might prefer to locally renumber the users to achieve a contiguous range of IDs rather than a scattering over a regular structure with subsequent remedies such as the MPT. 

However, before proceeding to our solution, we would like to evaluate the base case, the mask-controlled MT. We would like to establish some quantitative characteristics of MT paths under a random distribution of leaf occupancy.

\subsection{Sparse MT Statistics\label{sec:stats}}

\def\PDF{{\rm PDF}}

\begin{figure}
\begin{center}
\includegraphics[width=0.3\textwidth]{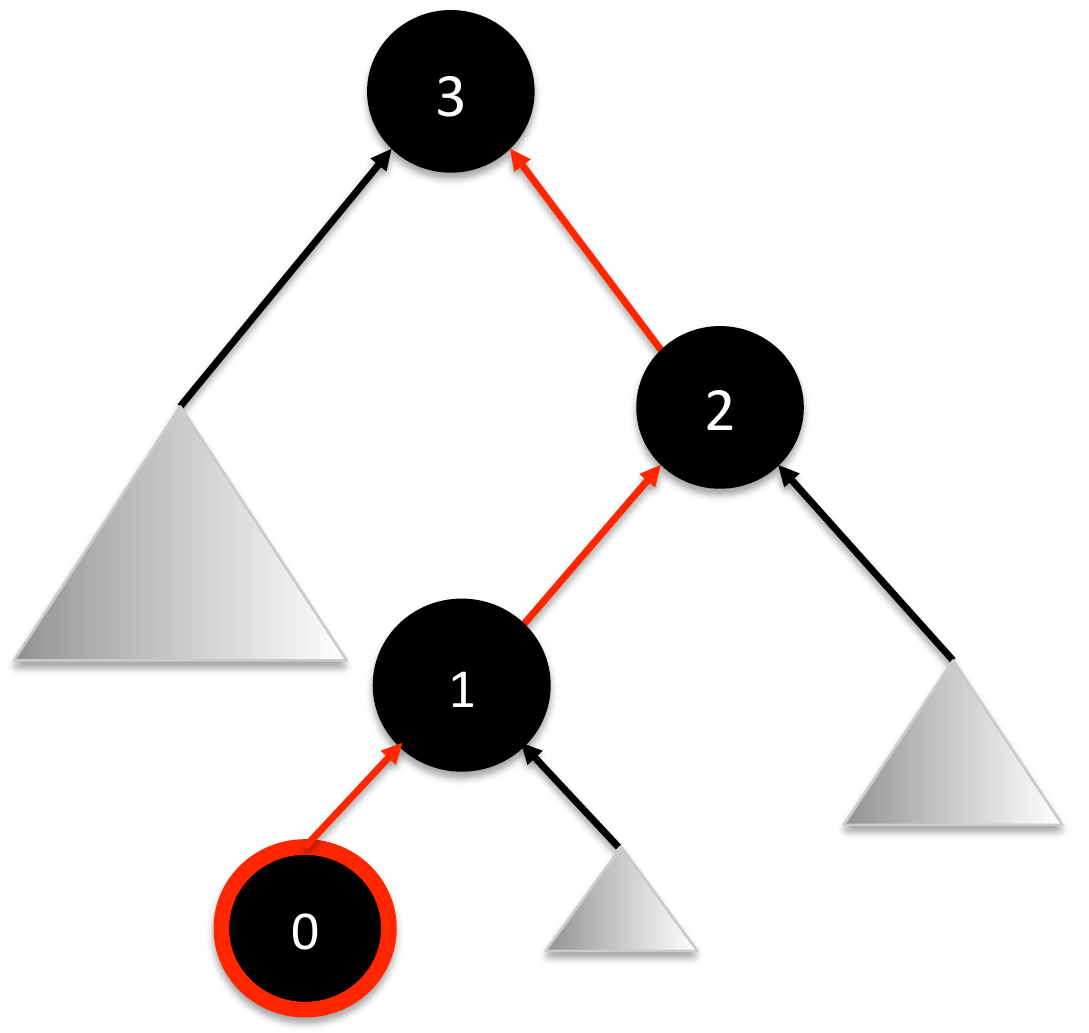}
\end{center}
\caption{A path across a sparse MT\label{fig:path}}
\end{figure}

Let us number the levels of the MT from the leaves up, starting with 0. We will call the number of adjunct hashes associated with a path its {\em weight}. Let function $\PDF_k(i)$ of integer $i$ be the probability for a path from a given leaf to a hight-$k$ node on the tree to be of weight $i$. Recall that we only count nodes on the path with both children being non-NULL as those require an adjunct hash for a Merkle proof. Let us take a closer look at an example path, see figure \ref{fig:path}. From level 0 we move to level 1 as a left child, then to level 2 as the left child, and finally to the level 3 as the right child. Clearly different leaves' paths differ by the choice of left- and right- ascension at each level, but the significance of the node does not depend on it: the node is only counted when {\em both} its children are non-NULL. The shaded triangles signify the subtrees that represent the other, {\em non-path} child of the corresponding path node, which, if non–NULL, produces what we termed above the adjunct hash. If the non-path node is NULL, this fact is noted in the path bit mask, but no adjunct hash is produced. 

At level 0, the subtree is of height 0 (it is a leaf), at level 1 it is of height 1 (connects two leaves), etc. 

\paragraph{Statistical model.\label{pg:stats}} We adopt a model relevant to the IoT case that the PLS blockchain was developed for. A \thing\ submits a message for inclusion in a block very infrequently. It does so at random with some small probability $p$, whose value depends on the duty cycle restriction, urgency of the sensor data and the available energy budget. Without loss of generality let us assume that $p\sim0.1$ in our examples, which will give us some intuition of what kind of figures may arise in practical work. This level of activity means that a \thing\ participates in roughly one block out of 10 or that about 1/10 of all blockchain users are active in any given PLS round. We also assume that the activities of different \things\ are uncorrelated, so any given leaf is either present or absent (NULL) irrespective of the presence/absence of other leaves. 

\paragraph{Path weight.} The subtrees in figure \ref{fig:path} will consequently be NULL-valued with the probability 
\begin{equation}
\alpha_k = (1-p)^{2^k}\,,\label{eq:alpha}
\end{equation}
where $k$ is the level at which the subtree is rooted. Let us introduce the Probability Distribution Function $\PDF_k(L)$ as the probability for the weight of a path from level $k$ to a leaf to be equal to $L$. Clearly $\PDF_k(L>k)=0$, and we also assume for convenience that for all $k$, $\PDF_k(L<0)=0$. It is easy to calculate $\PDF_1$ directly:
\begin{equation}
\PDF_1(0) = 1-p,\; \PDF_1(1) = p\,.\label{eq:rec0}
\end{equation}
Indeed, the other child of a given leaf of a height-1 tree is NULL with the probability $1-p$, producing no adjunct hash, so $L=0$ with that probability; otherwise (with the probability $p$) the other leaf is non-NULL, supplying a single adjunct hash. 

For a hight-$k$ path we have a combinatorial problem of calculating the probabilities of $2^k$ combinations of absence/presence of each adjunct hash (remember that these probabilities are completely independent as per our chosen statistical model). Instead of doing this, we observe the following recurrence relation between the paths to neighbouring levels:
\begin{equation}
\PDF_{k+1}(i)=\alpha_k\PDF_k(i)+(1-\alpha_k)\PDF_k(i-1)\label{eq:rec}
\end{equation}

Indeed, if the non-path child of the height-$(k+1)$ path node is NULL (this happens with the probability $\alpha_k$) the number of adjunct hashes that the path to height $k+1$ produces is the same as that to height $k$. Alternatively, if the non-path child is non-NULL, it produces one adjunct hash, and so the probability to produce L hashes for the whole path is the same as the probability to produce $L-1$ for the path to height $k$. The above equation is the weighted sum of those two outcomes, a mixture distribution. 

The significance of Eqs. \ref{eq:alpha}--\ref{eq:rec} is in the fact that they permit direct calculation of the PDF at any level above 1 very cheaply given the value of $p$. The PDF obtained can deliver various practical parameters: the average path weight:\[
\bar{L}_k = \sum_i i\times\PDF_k(i)\,,
\]
the standard deviation, the probability that a certain limit $L_{\rm max}$ is exceeded, etc., which are useful in designing bandwidth-limited communication protocols.

\begin{table}
\lstset{ 
  basicstyle=\footnotesize\ttfamily\bfseries\color{blue}        
}
\begin{center}
\begin{tabular}{c}
\begin{lstlisting}

 k   mean     0     1     2     3     4     5     6     7     8     9    10  
==============================================================================
 2  0.200   81.0  18.0   1.0
 3  0.390   65.6  30.0   4.2   0.2
 4  0.734   43.0  42.2  13.1   1.6   0.1
 5  1.303   18.5  42.7  29.7   8.1   0.9   0.0
 6  2.118    3.4  23.0  40.3  25.7   6.8   0.8   0.0
 7  3.084    0.1   4.1  23.6  39.8  25.0   6.6   0.7   0.0
 8  4.083    0.0   0.1   4.1  23.6  39.8  25.0   6.6   0.7   0.0
 9  5.083    0.0   0.0   0.1   4.1  23.6  39.8  25.0   6.6   0.7   0.0
10  6.083    0.0   0.0   0.0   0.1   4.1  23.6  39.8  25.0   6.6   0.7   0.0
\end{lstlisting}
\end{tabular}
\end{center}
\caption{Numerical evaluation of Eq \ref{eq:alpha}--\ref{eq:rec}. Probability Distribution Function (\%) of path weight vs height in a sparse MT ($p=0.1$)\label{tab:MT-stat}}
\end{table}

\begin{figure}
\begin{center}
\includegraphics[width=0.8\textwidth]{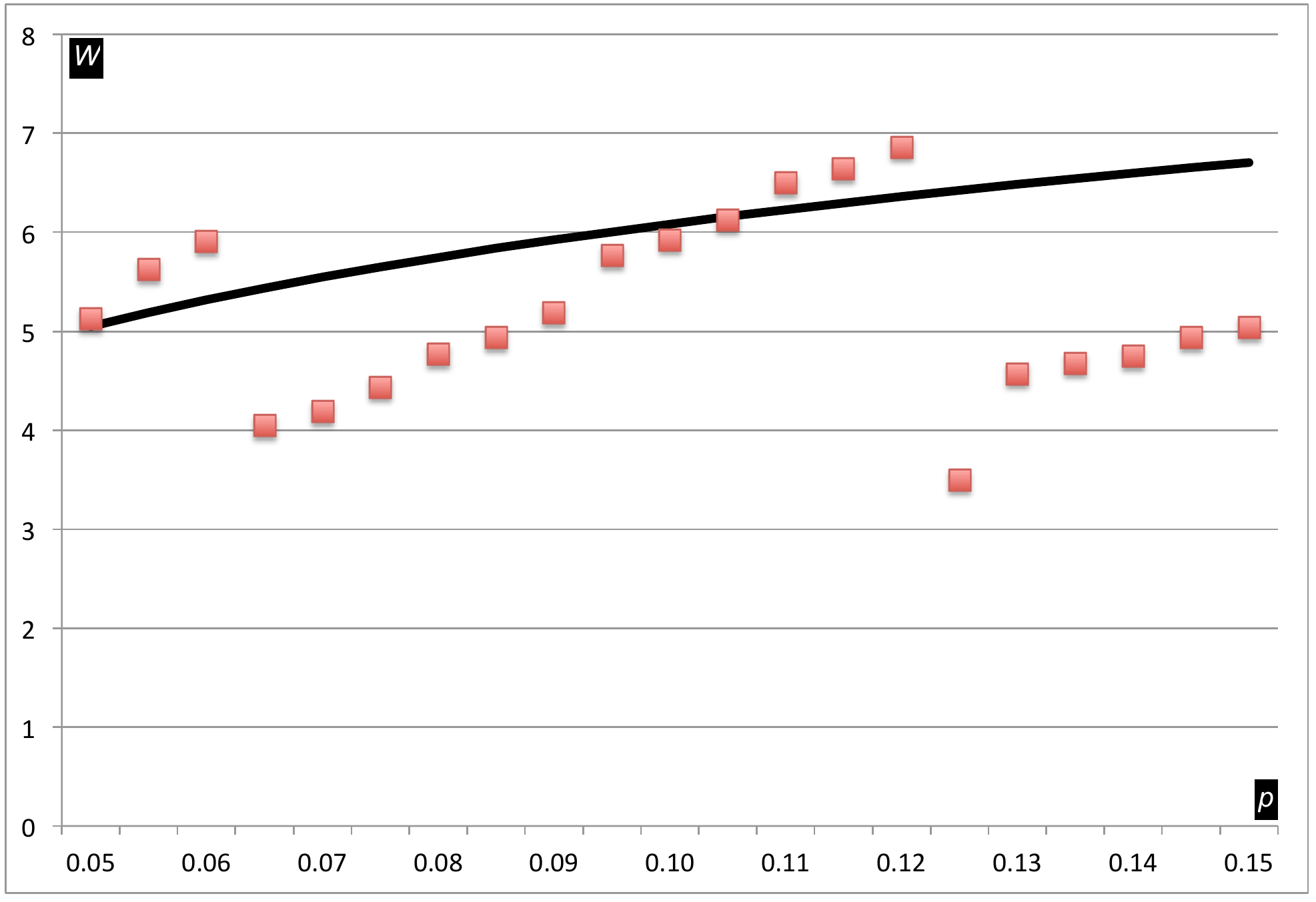}
\end{center}
\caption{Mean path weights. The curve: sparse MT, $n=1024$ leaves each occupied with probability $p$. Scattered dots: the average path weight of a truncated dense MT with $2^{\lceil \log_2 (np) \rceil}$ leaves\label{fig:path-stats}}
\end{figure}

Table \ref{tab:MT-stat} presents the outcome of a direct calculation of Eqs. \ref{eq:alpha}--\ref{eq:rec} for $p=0.1$ and also includes the value of $\bar{L}_k$ in the second column (heading ``mean''). The table shows the value (\%) of $\PDF_k(i)$, where $i$ runs horizontally. For obvious reasons nontrivial evolution only happens until $\alpha_k$ drops to small values, whereupon Eq. \ref{eq:rec} degenerates to
\[
\PDF_{k+1}(i)\approx\PDF_k(i-1)\,
\]
making the $\PDF(i)$ shift to the right by 1 without change of shape as $k$ increases. For $p=0.1$ sparsity is present in the first 7 levels of the tree; from level 8 up the tree becomes dense.

Another noteworthy feature of the distribution is its breadth: 95\% of the paths require from 4 to 8 hashes, with the mean being around 6, which would necessitate variable length communication, since a factor of 2 difference cannot be ignored. This variability comes despite the compression we have already applied by introducing the bitmap-controlled MT.

The sparse MT is indexed by the user ID, and a set of active users for an individual block is random as defined by our statistical model. To get a feel of how efficient the sparse MT is in terms of path weights, we compare its mean path weight with that of a truncated dense tree carrying the same number of non-NULL leaves. We use the least sufficient hight of the dense tree to accommodate all non-NULL leaves, and place all NULLs on the right hand side of level 0, so that the non-NULL leaves may be contiguous, and use Eqs: \ref{eq:none-case} to deal with none values (that is what we mean by truncation). Figure \ref{fig:path-stats} compares the path weights of the two trees. The dependence of the path-weight averaged across the truncated tree on the number of non-NULL leaves is not smooth, as the tree hight leaps up when the number of non-NULLs crosses power-of-two boundaries. Nevertheless, one can clearly see that the dense tree has a lighter path weight at the majority of probability values, becoming slightly worse just before the probability reaches a value that the equivalent dense tree must grow at.  

\paragraph{Challenge.} Now we are prepared to argue our case. A user engaging in the SLVP protocol as verifier must examine every consecutive block to see if the prover has placed a message in it (i.e. an S-, LV- or P-message). According to our statistical model, in a great majority of the blocs, in fact, in a factor of $1-p$ of them, the prover's message is likely to be absent. Nevertheless, the verifier needs to satisfy itself that indeed, no message from the prover is present. With the classic MT as well as MPT and our own version of sparse MT, the absence of a leaf is almost as expensive to prove as its presence with a particular value. The difference is that for an absent leaf the label is NULL and it is not communicated, but that is a difference of 1 against, as our calculation shows, circa 6 adjunct hashes to be communicated when 1024 users participate at probability $p=0.1$. This means that on average 60 (!) hashes would be required to certify the start of an SLVP round. Worse still, each active user, even when all it does is wait for a possible signed message from a counterparty, will be actively requesting the counterparty's root-adjunct path from CAS every time a block is released, which pretty much destroys the advantages of a low-bandwidth Guy Fawkes protocol. 

However, a simple remedy exists, which we consider next.

\section{Tunstall-Merkel Tree\label{sec:MTT}}

\paragraph{Basic idea.} We kill two birds with one stone by providing a one-time renumbering of users in each block while broadcasting the renumbering information together with the root of the tree. The purpose of the renumbering is to achieve a contiguous range of ID numbers. This way absent users will be recognised as such immediately by any counterparty involved. As a result the cost of absence proof will be zero (plus the cost of the one-for-all broadcast message, which need not be requested). The indexing structure in terms of new IDs will be the kind of tree we have already studied and shown the superior access cost of: a dense, truncated one.

How do we enumerate users that are present? Imagine a bitmap sized $2^h$, where $h$ is the height of the original (sparse) MT. In the bitmap 1s mark the presence of the corresponding user/leaf and 0s its absence. Under our statistical model (see page \pageref{pg:stats}) on average $2^kp$ bits (as per binomial distribution) of the bitmap will be 1s. Users are renumbered according to the bitmap: the user's new ID is the number of 1s in the bitmap preceding the bit that corresponds to the user's actual ID.

Our statistical model assumes that all users are engaged the whole time. A user that decides not to use the blockchain for a while will not be able to maintain a factor of $p$ messages per block on average until the user becomes active again; the user's bit position in the bitmap will be 0 during that period. If there are many such users, the bitmap may have significantly fewer 1s than the aforementioned expectation $2^kp$. In this sense the expectation is pessimistic.

The maximum number of users is within a near-unity factor from the number of \things\ in the swarm, since non-IoT users have typically a one-to-many relation with \things: a human or a server would be in control of several IoT devices. The majority of the users tend to be always-on, active \things, which work according to a near-periodic schedule. Another useful circumstance here is that the bitmap can be effectively and efficiently compressed to a fraction of its length, provided that the distribution of 1s is close to random and that the number of 1s is known to both the sender and the recipient. The former can be made true by pseudorandomisation, and the later is easy to achieve by including a small integer (typically 10-12 bits in length) in the message that broadcasts the bitmap. In this section we describe the compression technique, and in the next one we will propose a simple and efficient pseudorandomisation.

\begin{figure}
\begin{center}
\includegraphics[width=0.8\textwidth]{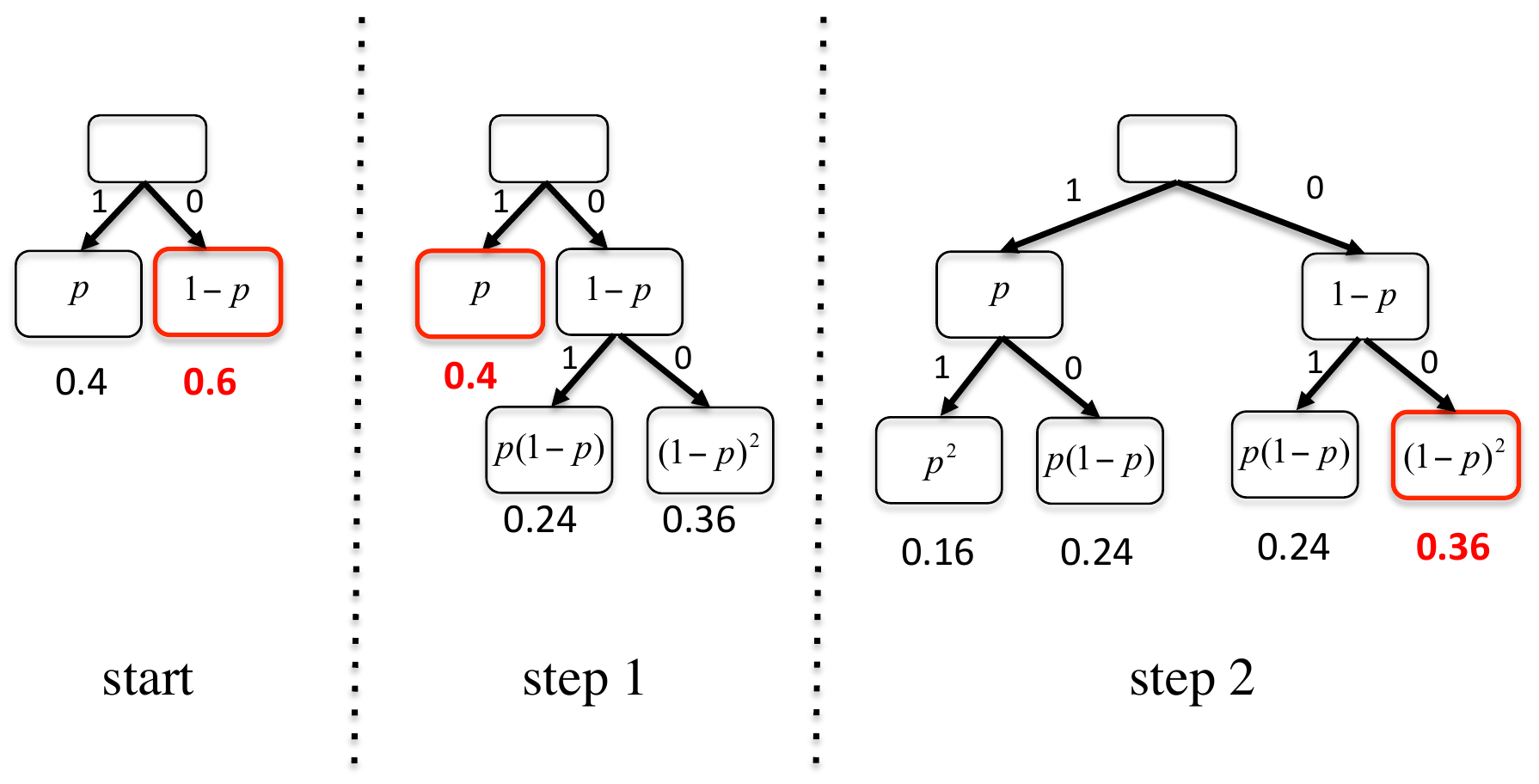}
\end{center}
\caption{Tunstall Tree for $p=0.4$. First two steps of the algorithm.\label{fig:t-tree}}
\end{figure}

\paragraph{Tunstall code.} Given a bit string of length $n$ which is expected to contain $m=pn, p<1$ ones in random positions (which makes $p$ a true probability), or alternatively a bit string which is known to contain $m$ ones, $m=pn$ in random positions (which makes $p$ an {\em empirical} probability), with the positions of ones pairwise uncorrelated (this is called a {\em zero-order} environment), we set ourselves the task of finding a bijective function $C:B^n\to B^{rw}$, $rw<n$ that maps the string to a sequence of $r$ codewords of length $w$. We wish to minimise $rw$, or, for a given $w$, to minimise $r$. The theoretical limit of compression is well known from information theory: $rw \ge H_0n$, where the zero-order per-bit entropy $H_0$ is defined thus:
\[
H_0 = -p\log_2p -(1-p)\log_2(1-p)\,.
\]
The mapping $C$ is realised by partitioning the source bit string into (generally unequal length) chunks and assigning a codeword to each. A chunk $b_0, b_1, \ldots, b_k$ is found at any given position in our random string with the likelihood 
\begin{equation}
p_c = \prod_{i=0}^k p^{b_i}(1-p)^{1-b_i}\,.\label{eq:lh}
\end{equation}
The best code with the word length $w$ should assign its $2^w$ codewords to the $2^w$ chunks with the highest likelihood. It must also make sure that the code is complete, i.e. any bit sequence can be represented as a sequence of codewords. It is intuitive that such a code would be optimal, and it can be proven that it is also asymptotically effective, i.e. that its compression ratio tends to the entropy limit as $w$ tends to infinity. 

It is not easy, however, to turn Eq. \ref{eq:lh} into a practical encoder/decoder. The main reason for it is that the value of $k$ is not bounded, and neither is the search for suitable chunks to find the top $2^w$ ones in terms of their likelihood. The problem is not so much the amount of work required for the search, since we could take the logarithm of Eq \ref{eq:lh} and maximise the linear form
\begin{equation}
\log p_c = l\log p + (k-l)\log(1-p)\,,\label{eq:lhlog}
\end{equation}
where \[
l=\sum_{i=0}^k b_i\,,
\]
in the $(0\le l\le k, k>0)$ area of the $(l,k)$-plane starting at the maximum $(0,1)$. The problem is that each $(l,k)$ point corresponds to ${k \choose l}$ chunks of length $k$ with $l$ 1s in each. Their enumeration and mapping at different $k$ would be rather awkward.

Tunstall in his PhD thesis\cite{Tunstall} proposed a greedy search which at the same time builds a compact dictionary structure (the Tunstall tree) that can be used for encoding/decoding efficiently, {\em without} sharing the dictionary (as long as $p$ is known to both the encoder and the decoder). The greedy search turns out to be of excellent quality, too, delivering the entropy limit asymptotically\cite{Tunstall}, and, as a recent study shows \cite{compression}, with a rapidly decreasing redundancy as $w$ increases. The redundancy formula from \cite{compression} being useful for our analysis, we present it below (without derivation and rewritten in our notation):
\begin{equation}
rw \le nH_0 + O(nH_0{\log(1/p)\over w})\,.\label{eq:redundancy}
\end{equation}

We will return to Eq \ref{eq:redundancy} later and present our own measurements for the relevant range of parameters, but let us first introduce the dictionary idea, see figure \ref{fig:t-tree}. The dictionary is a (generally imbalanced) labelled binary tree. Both the nodes and the edges are labelled. The edge labels are 0 and 1 as usual, and a node's label is the likelihood value of the chunk composed by reading the edge labels along the path from the root to the node. The algorithm builds the tree node-by-node, as follows:
\begin{enumerate}
\item Create a root node with two edges labelled 1 and 0 to two child nodes labelled with the value of $p$ and $1-p$, respectively.
\item Find and mark the maximum likelihood leaf node. Denote its label as $R$.
\item Create two leaf children of the marked node and connect them with edges labelled 1 and 0. Make the node labels the value of $Rp$ and $R(1-p)$, respectively.
\item Repeat steps 2 and 3 until the tree has $2^w$ nodes besides the root. 
\item Now relabel each leaf by its consecutive leaf number while visiting the leaves in some order agreed between the encoder and decoder\footnote{For example, left to right, or in prefix order. The properties of the code remain the same under any permutation of the codeword assignments but the practicalities of encoding/decoding require a shared order}. 

\end{enumerate}

Tunstall encoding is achieved by running the source bit-string down the Tunstall tree bit by bit until a leaf is reached, at which point the codeword is read off from the leaf label and the process returns to the root. Tunstall decoding requires a $2^w$-entry table where variable-length path sequences are set against codewords, with the former read off from the path to the leaf labelled by the latter. We notice with satisfaction that Tunstall decoding has the cost $O(1)$.

\paragraph{Implementation.} Tunstall encoding (and especially decoding) is very undemanding, well within reach of a small, system-on-chip smart sensor. To avoid accuracy/underflow problems with repeated multiplication in generating the dictionary at the receiver (Step 3 of the algorithm above), one could use log-likelihoods as node labels. Then instead of multiplication,  $\log p$ and $\log(1-p)$ are {\em added} to the parent label to produce labels for the 1- and 0-child, respectively\footnote{We use the fact that the greatest number has the greatest logarithm}. This way for any reasonable table size, computational accuracy will not be a problem. We implemented the algorithm to see what kind of residual redundancy we could be getting from a specific Tunstall code. The results of our running a Tunstall compressor through 1 million random bits are presented in table \ref{tab:tun-sim}. Comparing this with Eq  \ref{eq:redundancy}, we conclude that at $p=0.05$ the compressor already reaches asymptotic mode when doubling the codeword length roughly halves the redundancy $\rho$. At the same time, the dependency $\rho(p)$ in the interesting range of $p$, i.e. in the area around $p=0.1$ is not quite asymptotic: the contrast in redundancy between $p=0.05$ and $p=0.15$ is nowhere near a factor of 2 that the formula suggests. 

From the practical point of view, if we target a blockchain with  $\sim1K$ users, with 10\% of them posting messages in any given block, $p=0.1$ suggests a compressed bitmap of at least $1024\times0.47=482$ bits or 61 bytes. An 8\% residual redundancy would increase this by only 5 bytes. However, the bitmap is broadcast together with the root hash, 32 bytes long, and a few extra bytes of forced redundancy for the purposes of S-message verification (as per PLS protocol). This increases the length of the S-message up to nearly 100 bytes, and at this level a redundancy of the compressor to the tune of 5 to 10 bytes makes little difference. 

\begin{table}
\begin{center}
\begin{tabular}{ccccc}
$w$&$p$&$\kappa$&$H_0$&$\rho$(\%)\\
\hline
4&0.05&0.37& 0.29&30.4\\
4&0.10&0.50&0.47&7.5\\
4&0.15&0.65&0.61&6.9\\
8&0.05&0.32&0.29&13.4\\
8&0.10&0.49&0.47&4.6\\
8&0.15&0.63&0.61&3.8\\
\hline
\end{tabular}
\end{center}
\caption{Observed redundancy $\rho$ of Tunstall code. Sample length before compression: $10^6$. Column headers: $w$, codeword length; $p$, probability of 1; $\kappa$, compression ratio; $H_0$, per-bit entropy; $\rho=(\kappa-H)/H$, residual redundancy(\%)\label{tab:tun-sim}}
\end{table}

If the number of users drops to 0.05, even the poor compression quality for $w=4$ results in only 379 bits (though 82 bits, or 11 bytes, more than the entropy limit), which is still less than the already acceptable 482 bits we observed for $p=0.1$. An alternative is to use a list of raw ID numbers, circa 52 in total, each requiring 10 bits. This is 520 bits, far worse than the compressor's output, but not significantly worse than 482, and the list length would decrease in proportion to $p$. If $p$ were to drop further below 0.05, and if the Tunstall compressor further deteriorated, the uncompressed `list' option could at some point be preferred, with the switch controlled by a single additional bit in the message. 

\begin{table}
\begin{center}
\begin{tabular}{ccl}
codeword&$-\log_2 p_c$&chunk\\
\hline
0000 & 3.06 & 0000000000000 \\
0001 & 5.55 & 0000000000001 \\
0010 & 5.32 & 000000000001 \\
0011 & 5.08 & 00000000001 \\
0100 & 4.85 & 0000000001 \\
0101 & 4.61 & 000000001 \\
0110 & 4.38 & 00000001 \\
0111 & 4.14 & 0000001 \\
1000 & 3.91 & 000001 \\
1001 & 3.67 & 00001 \\
1010 & 3.44 & 0001 \\
1011 & 3.20 & 001 \\
1100 & 3.20 & 010 \\
1101 & 5.70 & 011 \\
1110 & 2.97 & 10 \\
1111 & 5.47 & 11 \\
\hline
\end{tabular}
\end{center}
\caption{4-bit Tunstall code for $p=0.15$ \label{tab:tun-code}}
\end{table}

We conclude that a four-bit Tunstall code is all that is required to implement the PLS S-message within half of the maximum LoRa message length (250 bytes). To aid the reader's intuition, we present an example of a 4-bit Tunstall code for $p=0.15$ in table \ref{tab:tun-code}. For each codeword we additionally show its log-likelihood. Notice that unless the log-likelihoods are exactly identical, as is the case for codewords 1010 and 1100 which correspond to chunks with the same number of 1s and 0s, the differences between log-likelihoods manifest themselves in the first (decimal) fractional digit  already, so computational accuracy should not be a concern\footnote{the main danger would be that the sender and the receiver use different floating-point arithmetic, incur different rounding errors and end up using different dictionaries; this example shows that for a small dictionary there is no such danger}.  

\paragraph{New structure of the root of trust.} In the original PLS protocol \cite{PLS} the S-record was a message that contains the block's root of trust $J_i$, which was the root hash of the Merkle tree representing the new block $B_i$. In the light of our analysis of indexing costs presented in Section \ref{sec:stats} and the properties of Tunstall encoding described in the current section, we propose to modify the root of trust $J_i$ as shown in Table \ref{tab:root}.
\begin{table}
\begin{center}
\sf\small
\begin{tabular}{cccl}
\hline
offset & field & size& description\\
(bits) &  & (bits) & \\
\hline
0& $T_i$ & 256 & root hash of the Merkle Tree for the new block $B_i$\\
		 &&& built using new user-IDs\\
256 & $n$,$m$ & 24 & $n$: total number of users, $m$: how many present \\
280 & flags $\phi$& 8 & bits 0,1: bitmap type (plain, compressed, list, empty) \\
			  &&& bit    2:  (0: $w=4$, 1: $w=8$)\\
			  &&& bits 3--7: pre-randomisation parameter (see next section)\\
288 & bitmap $M$& $L\le1024$& processed bitmap content  \\
$L+288$& redundancy& 32 & all zeros, for PLS validation \\
\hline
\end{tabular}
\end{center}
\caption{Structure of the proposed block root-of-trust $J_i$\label{tab:root}}
\end{table}

The total message length is $L+320$ bits or $L+40$ bytes. We expect $L$ to be close to 60 bytes in most cases (which is the entropy limit for 1024 users at 10\% occupancy per block on average), which makes the S-message circa 100 bytes long, but if necessary $L$ can be increased to 128 bytes resulting in the packet length 168 bytes, still well within the length limit (255 bytes) for LoRa communications. A 128-byte bitmap would support the number of users up to 1024 without Tunstall compression, or about twice as many if Tunstall compression is used at 10\% occupancy.

The hash $T$ requires the server to renumber the users, building a new Merkle tree and computing its root hash. The client will recompute $T$ from any leaf hash, adjunct hashes and the path mask -- all sent to it by CAS (unauthenticated, unsigned) at request, and then check it against the $T$ in the S-record. 

Notice that the redundancy field is only 32 bits, since it is impossible to crack the S-message directly: {\em both} the plaintext and the key are unknown, the former due to the XORing of the next P-message, yet to be received, to the plain text, see Figure \ref{fig:pls-prot}. As mentioned earlier, the purpose of the redundancy field is to thwart a {\em random message} attack for the DoS purposes, and so a 32-bit redundancy translates into a less than 1-in-a-billion chance to cause the recipient to accept a false message, which is more than sufficient in the IoT world.

Finally, let us dwell a little on the block's MT whose root $T_i$ is included in the block's root-of-trust $J_i$. The leaves of that tree are hashes of the user records with the user ID corresponding to the path label sequence as usual, except the IDs are now new IDs calculated from the block bitmap and occupying a range from 0 to $m-1$ without gaps. Since $m$ is not necessarily a power of 2, the MT generally consists of a complete half with leaf labels in the interval $[0, 2^{\lfloor\log_2 m \rfloor})$ without gaps and a truncated half with labels in the interval $[2^{\lfloor\log_2 m\rfloor}, m)$, also without gaps, with the rest of the leaves labelled with NULL. The shape of the MT depends solely on one parameter, $m$, which is part of the root-of-trust. Consequently, no further information, such as path masks, etc, is required for access and validation of the root hash, $T_i$, except, of course, we must use Eqs. \ref{eq:none-case} to calculate the root-adjunct paths in the truncated half. For our running example of 10\% occupancy and the total number of users 1024, the value of $m$ will have an expectancy of around 102, which means this path will be between 1 and 7, and never longer. Table \ref{tab:MT-stat} indicates that the standard (MPT or mask-controlled MP) would require from 3 to 9 adjunct hashes. The difference between 7 and 9 is not big, but notice that 8 hashes would already require more than one LoRa packet to transmit. 

We would like to emphasise here that the main effect of using the Tunstall-Merkle Tree (TMT, which is how we wish to call our construction) rather than, say, MPT is {\bf not} that fewer hashes have to be communicated with the former than the latter, but the fact that the latter requires a full path irrespective of the presence or absence of the leaf for secure retrieval. By contrast, a TMT provides an absence proof directly from the root-of-trust bypassing the Merkle Tree entirely. Since the SLVP protocol requires every \thing\ to check the presence of its own S- and LV- messages before advancing the protocol, and since a \thing\ would typically monitor another user's infrequent activity, the cost of absence proof dominates over the cost of secure retrieval. Nevertheless, it is reassuring to see that the latter is also improved, in terms of limits if not necessarily average, by our approach. The price we are paying is some additional calculations well within the capabilities of resource-limited systems such as most \things\ tend to be.

Returning to the compression issue, there is one factor yet to be accounted for. We remarked earlier that our statistical calculations are based on the zero-order assumption, i.e. that different users' behaviours are uncorrelated. Obviously it is not the case when users engage in a higher-level protocol with one another, e.g. producer/consumer. This may skew the chunk statistics resulting in a longer codeword sequence for the block bitmap. In the next section we will propose a simple remedy.

\section{Pre-randomisation\label{sec:rand}}

The idea is to apply a bijective function to the source user ID which depends on an extra parameter, block number $i$. A different block number should result in a very different permutation. This way a position in the block bitmap will have the value 1 in a proportion of bitmaps that does not depend on the value of other positions. A user ID will be associated with a pseudorandom sequence of positions as new blocks are produced. Invertibility (bijection) is very important, as it prevents different users from being mapped on the same bit-position in the bitmap, thus ensuring that the mapping a pseudorandom permutation. 

A simple and effective pseudorandom permutation based solely on the block number $i$ can be achieved by analogy with randomising the order in a deck of playing cards. One player performs the deck shuffle: for the card in position $i$ in the deck $i\in[0,n=2^d)$ represented in binary as $i=i_di_{d-1}\ldots i_0$, the new position of the card $i^\prime$, represented in binary as $i_d^\prime i_{d-1}^\prime\ldots i_0^\prime$ is obtained from the current position by applying the following operator
\[
i_d^\prime i_{d-1}^\prime\ldots i_0^\prime = i_{d-1}i_{d-2}\ldots i_0i_d = \sigma(i) \,,
\] 

which is a cyclic shift left. This corresponds to dividing the deck into two halves and interleaving them, exactly as an experienced dealer would.

When the shuffle is finished, the other player {\em shifts} the deck, i.e. divides it into two unequal parts and transposes them. In terms of card numbers, this corresponds to adding a pseudorandom value $v$ modulo $2^d$:
\[
i^\prime = i + v  \mod n = \tau_v(i) 
\] 
Applying the shuffle-shift operator $Q_v = \sigma \circ \tau_v$ to a range $\rho=[0,n)$ $t$ times with a pseudorandom choice of $v$:
\[
Q_{v_t} \ldots Q_{v_1}Q_{v_0}\,\rho\,
\]
has the same effect as repeatedly shuffling/shifting a deck of cards, which, the intuition suggests, delivers a rather arbitrary permutation. We call $t$ the number of rounds. 

Note that the operators $\sigma$ and $\tau_v$ have a negligible cost even when executed by the least powerful platform, as they take literally a few machine instructions. The cost of pseudorandom generator that produces a series of $v$-values is similarly small if we use a standard Linear Congruential Generator (LCG): 
\[
v_{k+1} = F\times v_k+1 \mod n,\;0\le k<t\,,
\]
where $n$ is a power of 2, $v_0$ is set to the block number, and the factor $F$ is any positive integer that satisfies the well-known Hull-Dobell constraint: $F= 5 \mod 8 $. We chose for $F$ the hex value {\tt 5EED} which satisfies the constraint and which has more than enough significant digits for any reasonable $n$.      

Our solution appears quite attractive from the point of view of its cost; however, while bijectiveness is guaranteed by construction, we need to be reassured that the solution can deliver sufficient {\em randomness} of mapping at a reasonably small number of rounds $t$. 

\paragraph{Avalanche test.} How do we judge the quality of a pseudorandom mapping? A common test is based on the so-called avalanche criterion\cite{avalanche}, used in evaluation of symmetric ciphers and hash functions. We consider it next in relation to our mapping $Q$.

Select the block-number $v_0$ for the test.  Next, select a number $x$ from the range $\rho$ and some integer $0\le k<d$ and prepare two numbers $x_1=x$ and $x_2$ same as $x_1$, except bit $k$ of it is flipped. Apply the mapping $Q$ to both and take the bitwise XOR of the results: $A_i(x,k) = Q(x_1)\oplus Q(x_1)$. Let $A_i(x,k,l)$ be the $l$th bit of $A_i(x,k)$. Define the correlation matrix ${\bf K}_{kl}$ thus:
 
\begin{equation}
{\bf K}_{kl} = \langle A_i(x,k,l) \rangle_{x,i}\,,\label{eq:corr}
\end{equation}

Here the averaging is done over values of $x\in\rho$ and block numbers $v_0$. Good randomness of the mapping manifests itself in the closeness of all matrix elements of ${\bf K}_{kl}$ to 1/2:
\begin{equation}
\max_{kl} ({\bf K}_{kl}-1/2) \ll 1/2\,.\label{eq:avalanche}
\end{equation}

This means that if we flip a random bit in a random value $x$ the probability that any bit in the image of $x$ under $Q$ flips in response is close to 1/2. In other words, if we flip one bit in $x$, on average close to one half of the bits in the result will flip. The name ``avalanche effect'' is to do with the fact that small changes cascade through the rounds of the computation causing further changes until all bits of the result are affected in a complex and unpredictable, though of course deterministic, way. 

We have applied the avalanche criterion to our proposed randomiser to estimate the acceptable minimum value of $t$, the number of rounds. As it is impossible to average over all potential values of the block number $v_0$, we limited ourselves to 50 random samples taken from the interval $[0,10000]$ The results have proven quite insensitive to the averaging over $v_0$, which in not surprising given that we established that the required $t$-numbers are in the hundreds. The averaging over $x$ was done by sweeping the whole range $\rho$.

\begin{table}
\begin{center}
\begin{tabular}{ccl|ccl|ccl}
$d$& $t$& $\delta$&$d$& $t$& $\delta$& $d$& $t$& $\delta$\\
\hline
10&100&0.061&11&100&0.113&12&100&0.164\\
10&150&0.014&11&150&0.026&12&150&0.042\\
10&200&0.008&11&200&0.009&12&200&0.013\\
\end{tabular}
\end{center}
\caption{Avalanche test of the pre-randomiser. $d$: input length (bits), $t$: number of rounds, $\delta$: mapping quality, $\delta=\max_{kl} ({\bf K}_{kl}-1/2)$, see Eq \ref{eq:avalanche}\label{tab:avalanche}}
\end{table}
 
The results of the avalanche test are presented in Table \ref{tab:avalanche}. For practical purposes we limited ourselves to $n=1024$, 2048 and 4096, since more users are unlikely to be supported by the communication infrastructure of a single site. The results show that a surprisingly large number, around 200, of rounds is required to achieve good randomisation. It is large compared to the number of rounds one expects to be necessary to randomise a deck of cards (of the order of 10), but it is not large technically: a microprocessor would have to execute only a few thousand instructions to compute the image $i^\prime$ given $i$. This has to be done as many times per block as the number of counterparties that the user has to monitor on the blockchain. For a \thing\ this would be a number of the order unity, hence the cost would be negligible even on a tight energy budget. On the other hand, setting $t$ to 200 would ensure that possible correlations between bits in the image do not exceed 3\% (0.013 normalised by 1/2), which should be good enough for practical purposes. 

Finally, let us recall that the IoT platform is low-power, but the server running the PLS protocol via the Sequencer is not. It has ample capacity to analyse the quality of the permutation in terms of its influence on Tunstall compression. We have reserved 5 bits in $\phi$ to pass to the shuffle-shifter an integer value in the interval $[0,32)$. It is convenient to use the value 0 to indicate that a random permutation is not required\footnote{This could be advantageous when, for example, no compression is used.}, whereas a nonzero value is added to the round counter $t$. The server can try up to 31 additional rounds and choose the one that gives the best compression.  The users receiving the root of trust will be aware of how many additional rounds should be performed and will maintain consistency.

\section{Putting it all together\label{sec:together}}

\begin{figure}
\begin{center}
\includegraphics[width=0.8\textwidth]{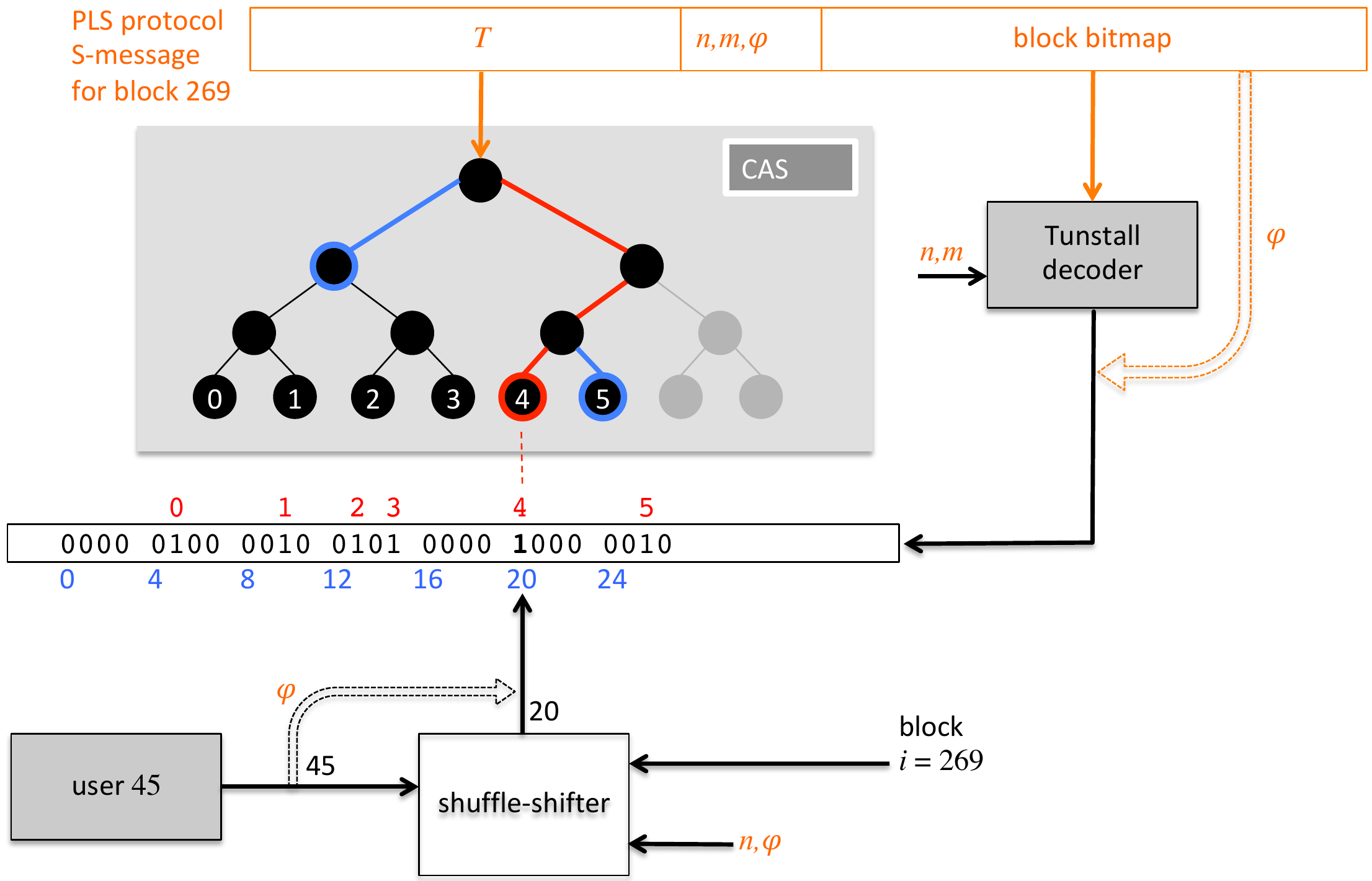}
\end{center}
\caption{Retrieving a user record from block 269 for ID$=45$; for this block $m=5$. \label{fig:overall-1}} 
\end{figure}

\begin{figure}
\begin{center}
\includegraphics[width=0.8\textwidth]{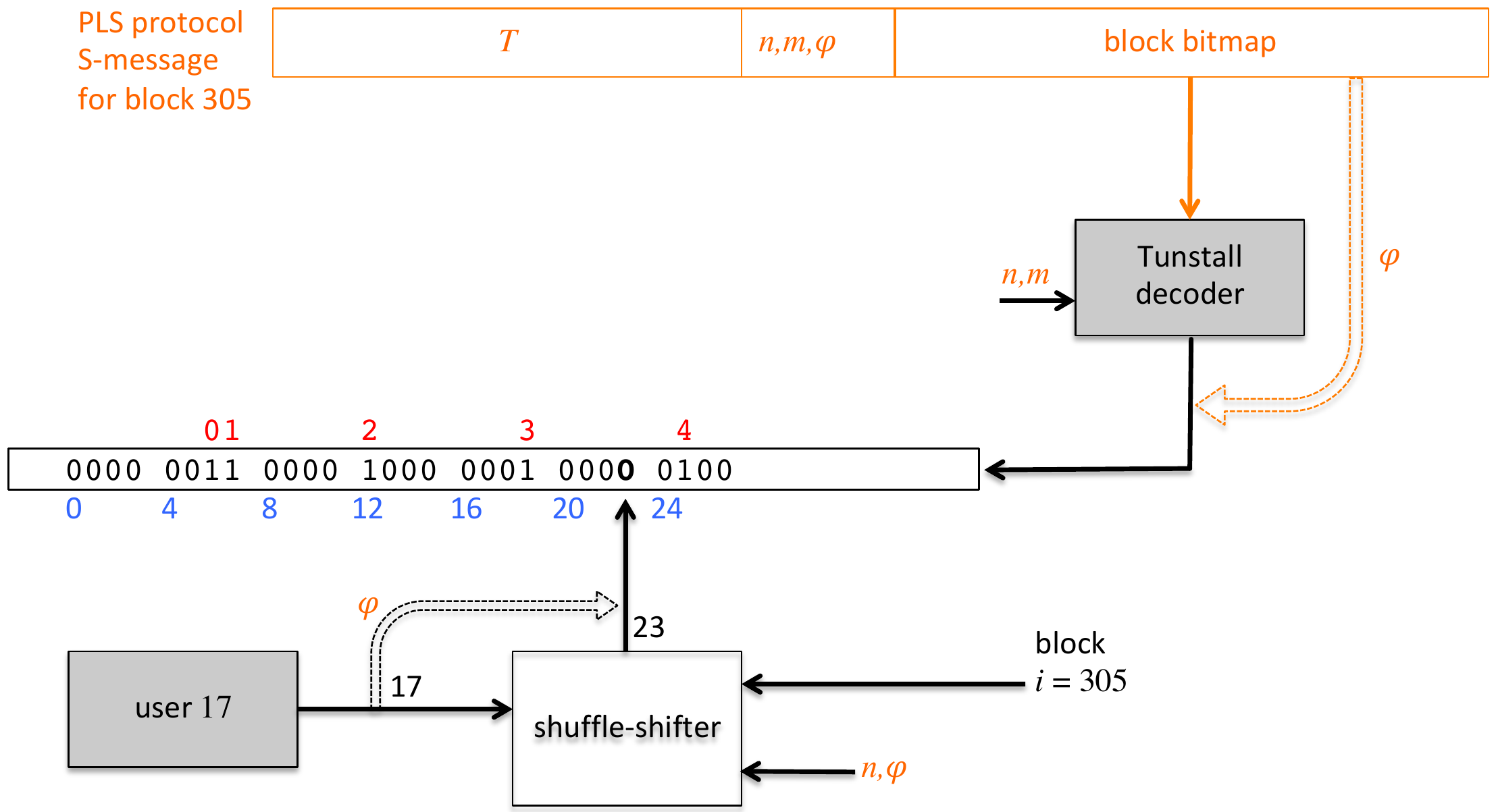}
\end{center}
\caption{Obtaining proof of absence.\label{fig:overall-0}}
\end{figure}

Next we consider a complete example of a user attempting to retrieve a contribution to a block that has been made either by itself or a counterparty. 

Figure \ref{fig:overall-1} presents the flow of data when a block-269 S-message is received and successfully unlocked by a user. The user is about to request the contribution to block 269 from user ID 45. To accomplish this, it needs to decode the received block bitmap by feeding it to the Tunstall decoder together with the parameters $n$ and $m$ (total number of users and the number of users contributing to block 269, respectively). The decoder produces the uncompressed bitmap. At the same time the user ID (45) and the block number (269) along with the total number of users and configuration parameters are fed to the shuffle-shifter, which will extract the number of additional rounds from $\phi$ and produce its output value, 20. The bit in position 20 of the uncompressed bitmap happens to be 1, which means that the contribution from ID 45 is present in block 269. The number of 1s in the bitmap to the left of position 20 is 4, so the index in the truncated Merkle tree for the contribution in question will be 4. The path to leaf 4 is highlighted in red in the figure.   

The user's CAS request will include the block number, 269, and   the ID index, 4. CAS will respond with the leaf hash $h_4$ and the adjunct sequence
\[
V_0=h_5,\; V_2=H(H(h_0\parallel h_1)\parallel H(h_2\parallel h_3))\,,
\]
which consists of the labels of the two nodes of the tree marked in blue. Because the user acquired $m$ from the unlocked S-message, i.e. the root of trust, it knows the shape of the tree. Consequently, no mask is communicated, but the user is able to reconstruct the mask anyway. 

To validate the requested $h_4$, the user checks that the following equation holds: \[
H(V_2\parallel H(h_4 \parallel V_0)\parallel H(h_4 \parallel V_0)^\prime)) = T\,.
\]

where $T$ is the root hash received with the unlocked S-message.

An alternative scenario is shown in figure \ref{fig:overall-0}. When attempting to retrieve the contribution of user ID=17 to block 305, it turns out that the output of the shuffle-shifter points to a 0 in the uncompressed block bitmap. Since the unlocked S-message is the root of trust, this constitutes a proof that block 305 has no contribution from user 17. Notice that CAS is not involved in the process at all.

\section{Related work}

The PLS blockchain and the protocols in basic form were proposed in \cite{PLS}. The idea of sparse Merkle Tree has an unclear origin. To the best of our knowledge it was first put forward by Bauer\cite{SMT-orig} and was recently improved on in \cite{SMT-rw}. Both studies are concerned with mutable trees, with objectives very different from ours, although, like ourselves, the authors remark on the importance of proofs of absence (non-membership). Tree statistics is tackled theoretically in \cite{MT-stats} in the context of optimising mutable MTs for the Bitcoin blockchain in the context of Bitcoin transactions. The objectives of this study are similar to ours as the authors attempt to group the leaves together to minimise the proof length, but they do it using tree transformations (taking the data structure red-black tree as a starting point), while we achieve a similar objective by renumbering the keys (user IDs in our case). 

The compression technique we use is due to Tunstall \cite{Tunstall} and this seems to be uniquely suitable for our case since it is based on empyrical probability of leaf occupancy, which is available to the Fog Server running PLS and which takes next to no resources to communicate to the client. The efficiency of our technique depends on this method.

We used our own pseudorandom permutation as a combination of a perfect shuffle and a random shift, using a classical LCG source \cite{Lehm51}.  There exist various methods of pseudorandom permutation, an oft-cited one being Fisher-Yates shuffle\cite{fisher-yates}, first published in the 1930s (citation unavailable). The idea there is to choose a (pseudo)random element of a sequence of source items and exchange it with the first element on the sequence. Clearly, if this is repeated enough times then any possible permutation could be achieved\footnote{unless the pseudorandom generator producing the selections is caught in a cycle first}. A recent paper \cite{mergeshuffle} presents a fast, parallel algorithm that mimics the technique of merge-sort except the merge makes a pseudorandom choice when ordering two elements for the output. 

However, our situation is quite different. Not because we are dealing with a contiguous range of numbers rather than an abstract sequence of objects: one could enumerate the objects and the problem would boil down to the one we are faced with. Our situation is different because the sender and the recipient must choose the {\em same} permutation. To encode an arbitrary permutation of $n$ numbers would take close to $(m-1)\log_2m$ bits, which is the same order of magnitude as the block bitmap we are trying to make more compact. Of course the ability to perform an arbitrary permutation is not required: all we want is break correlations between user IDs in a series of block bitmaps, and for this any sufficiently rich subgroup would do. 

\section*{Conclusions}

Statistical analysis of a sparse Merkel Tree under the assumption of uniform, uncorrelated leaf occupancy has been presented. The model obtained allows direct computation of the Probability Distribution Function for path lengths given the leaf-value probabilities. The path weight was quantified in terms of the number of adjunct hashes required for its leaf proof. We determined that the mean path weight of a sparse MT tree is close to that of a dense, truncated MT tree, with the latter being slightly better at most leaf-probability values $p$ in the practically interesting interval. We proposed an alternative structure, a Tunstall-Merkle tree, which combines a dense, truncated MT and a Tunstall-compressed bitmap indicating leaf ocupancy. We tested the compressor at several practical values of code size and quantified its residual redundancy. We found that a very small code table (16 or 256 codewords) proves sufficient for achieving near-limit compression, which means that Tunstall decoding presents no storage problem whatsoever to an IoT platform. To improve the effect of compression we further proposed a decorellation facility in the form of a shuffle-shifting algorithm and tested its properties using the standard avalanche criterion to determine the number of rounds. Both the Tunstall decoder and the shuffle-shifter with the codes size and the number of rounds, respectively, sufficient for our purposes are quite processor-efficient as well, since they involve inexpensive operations (table indexing, cyclic shift and binary addition) and short instruction sequences in implementation.

The main effect of the proposed technology is a drastic improvement in the cost of the SLVP protocol. Indeed an SLVP verifier has to check every block for the presence of counterparty (prover) contributions, and no such contribution would be present in a great majority of blocks. Our proposed Tunstall-Merkle tree has zero proof-of-absence cost, and when a leaf is present the communication cost of retrieval is in most cases better than that for the standard MT and MPT. Obviously our technique offers no advantage to a system with an unlimited and dynamic number of users, but it is beneficial for at least the PLS blockchain situation. The statistical analysis of a sparse MT/MPT has significance beyond the area of our study; it could be useful for planning and designing any secure storage structure that involves Merkle trees.

Future work will concentrate on higher-level protocols which control the interaction of \things\ with a smart contract within the same limited-resource set of assumptions.

\bibliographystyle{plain}
\bibliography{GF-BC}
\end{document}